\documentclass[prx,preprint,superscriptaddress,amsmath,amssymb,floatfix]{revtex4-1}
\usepackage[latin1]{inputenc}
\usepackage{graphicx}
\usepackage{float}
\usepackage{amsmath}
\usepackage{natbib}
\usepackage{esint}
\usepackage{eucal}
\usepackage{color}

\begin{document}

\title{Ultrasonic nodal chains in topological granular metamaterials}

\author{Aurelien Merkel}
\affiliation{Department of Physics, Universidad Carlos III de Madrid, ES-28916 Legan\`es, Madrid, Spain}
\author{Johan Christensen}
\affiliation{Department of Physics, Universidad Carlos III de Madrid, ES-28916 Legan\`es, Madrid, Spain}

\begin{abstract}
Three-dimensional (3D) Weyl and Dirac semimetals garner considerable attention in condensed matter physics due to the exploration of entirely new topological phases and related unconventional surface states. Nodal line and ring semimetals on the other hand can facilitate 3D band crossings characterized by nontrivial links such as coupled chains and knots that are protected by the underlying crystal symmetry. Experimental complexities, detrimental effects of the spin-orbit interaction, along with the merging of the underlying surface states into the bulk pose great challenges for growing advancements, but fortunately enable other systems, such as bosonic lattices, as ideal platforms to overcome these obstacles. Here we demonstrate a 3D mechanical metamaterial made of granular beads, which is predicted to provide multiple intersecting nodal rings in the ultrasonic regime. By unveiling these yet unseen classical topological phases, we discuss the resilience of the associated novel surface states that appear entirely unaffected to the type of crystal termination, making them a superb platform in ultrasonic devices for non-destructive testing and material characterization.
\end{abstract}

\maketitle

Topological phases of matter in insulators and superconductors have recently been extended to semimetals, thus broadening the family of exotic topological states \cite{RevModPhys.83.1057,RevModPhys.90.015001}. This active frontier in condensed matter physics recently explored topologically protected degeneracies in Dirac and Weyl semimetals that are identified by topologically robust band-touching manifolds. These unconventional fermionic semimetals are characterized by nontrivial band touching in the form of zero-dimensional discrete points, one-dimensional nodal rings and lines, or two-dimensional nodal surfaces. Peculiar topologically protected $\textbf{k}$-space geometrical manifolds appear in various interlaced nodal chains and links whose nontrivial linking carry a toroidal Berry phase of $\pi$, provided the loops enclose the aforementioned nodal shapes \cite{PhysRevB.84.235126,NatureNodal}. 

The nontrivial Berry flux is responsible for the formation of topologically protected surface states that emerge from the intersecting points of nodal shapes. E.g., bulk Weyl points give rise to surface Fermi arcs, whereas nodal rings host flat drumhead surface states that are relevant for topologically robust transport. Various materials have been predicted to sustain Dirac or Weyl fermions and contemporary experiments report on the observation of anisotropic or negative magnetoresistance in three dimensional semimetals \cite{PhysRevX.5.031023,Dirac1,Dirac2,AEM,Hong_2018}.

As opposed to nodal points and lines, nodal rings can generate exotic intersecting formations in the form of chains, knots and Hopf links that opens new horizons for unprecedented topological properties, unusual surface states and novel physics in general. A vast amount of theory has been devoted to this frontier in Heusler compounds, alkaline earth metals and electrides \cite{TheoNatComm,PhysRevB.96.041102,PhysRevB.96.041103,PhysRevB.96.201305,PhysRevLett.119.156401,PhysRevLett.120.106403,PhysRevX.8.031067}, whereupon experimental verifications of nodal rings have already been observed in non-centrosymmetric superconducting compounds and Zirconium based structures \cite{NatcommHasan,PhysRevB.93.201104,PhysRevLett.117.016602,NPJQuantumMaterials}.  

Bosonic settings such as photonic, phononic and sonic crystals are widely studied with the aim to tailor classical wave properties. Accordingly, given the complexity of electronic systems to unveil Weyl and Dirac semimetal physics as well as those peculiar intersecting nodal rings, man-made bosonic system have become increasingly popular to explore in the waves based context, the existence of Dirac-like plasmons \cite{PhysRevLett.110.106801}, optical Weyl points and Fermi arcs \cite{Lu622,NatPhysRectsmann}, hourglass nodal lines \cite{PhysRevLett.122.103903}, metallic mesh nodal chains \cite{NatPhotoNodal} and acoustic Weyl points \cite{MX1,FengLi1} including topologically protected negative refraction \cite{NatureZhengyou}. \\

\begin{figure}[ht!]
\begin{center}
\includegraphics[width=1.0\columnwidth]{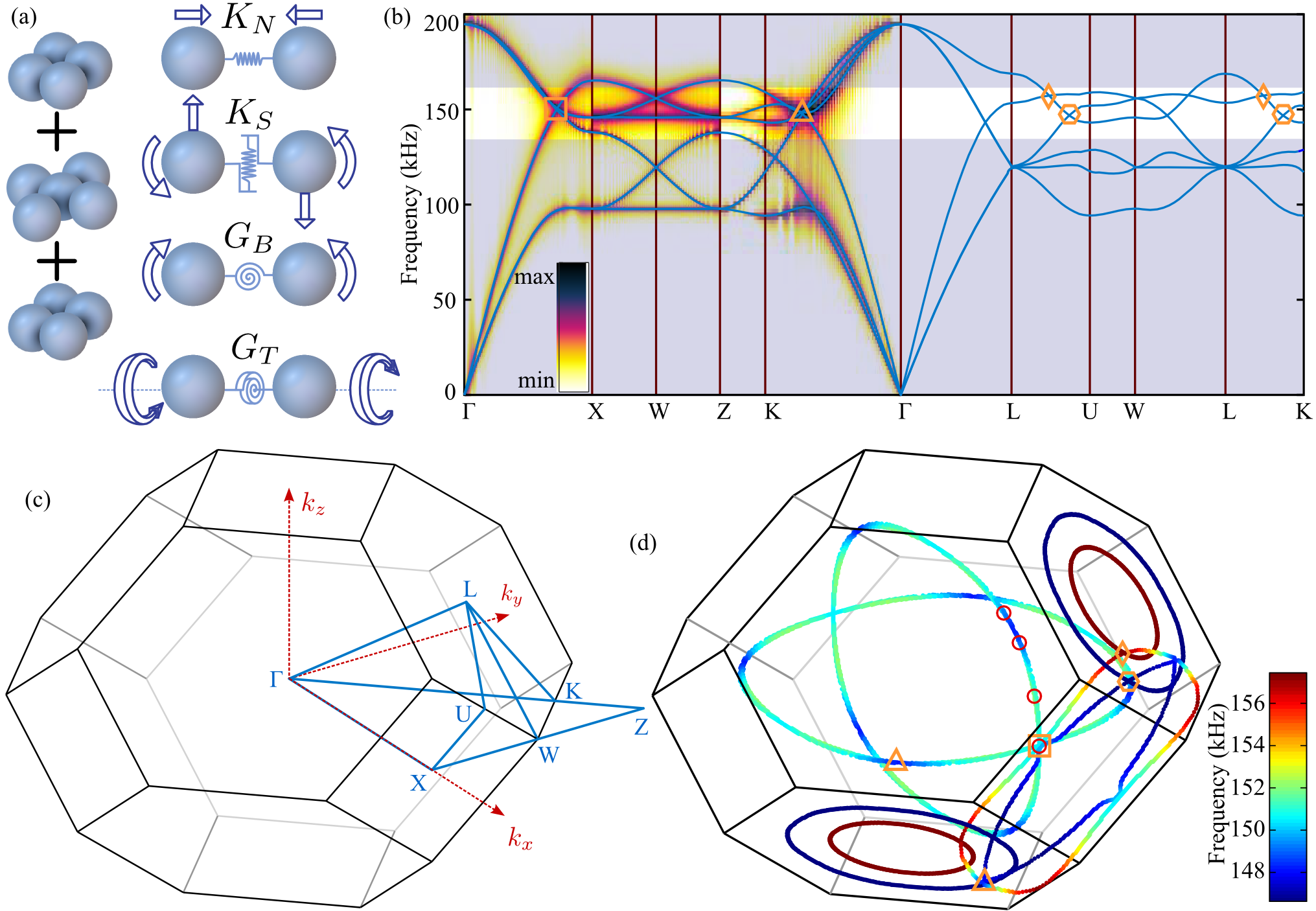}
\caption{Band structure of the FCC granular metamaterial. (a) Schematic of the structure comprising four different types of interactions between the beads. (b) Analytical band structure over the entire Brillouin zone. On the same diagram we numerically compute the dispersion relation for a periodic superlattice (see Methods). (c) The first Brillouin zone of the FCC crystal. (d) The nodal chain structure in momentum space evaluated in the ultrasonic frequency regime highlighted in (b). The orange squared, triangles, hexagons and diamonds mark the crossing points from the band diagram in (b). The red circles show the positions of the nodal lines that are discussed in Fig. \ref{figsurface}.  \label{figband}}
\end{center}
\end{figure}

In this work, we present an excerpt of the catalogue of nodal-chain semimetals for classical mechanical waves in man-made granular metamaterials. By solving the equations of motion for these artificial discrete media, we are able to engineer exotic mechanical intersecting nodal rings whose surface states at the truncated lattice interfaces display a remarkable durability. The elastic behavior of granular media drastically differs from conventional elastic waves in solids. At frequencies below the first spheroidal resonance of  individual beads, the structure is modeled by rigid masses of finite size connected by stiffnesses originating from the contact laws between the beads, thus forming a discrete crystalline lattice. Specifically here, we consider a face centered cubic (FCC) arrangement as illustrated in Fig. \ref{figband}(a). In the general case in the linear approximation, the contact between two grains is described by one shear stiffness $K_S$, one torsion stiffness $G_T$ and one bending stiffness $G_B$ accounting for the sliding, twisting and rolling resistances at the level of the contact, respectively, in addition to the usual normal stiffness $K_N$ as sketched in Fig. \ref{figband}(a). As a consequence, the rotational degrees of freedom of each individual particle play an important role in the dynamics of granular media. 
The linear equations of motion for one bead $\alpha$, with its infinitesimal displacement $\textbf{u}^{\alpha}$ and angular rotation $\textbf{w}^{\beta}$ around its equilibrium position in a monodisperse granular assembly, read for translation
\begin{equation}
m_b \frac{\partial^2 \textbf{u}^{\alpha}}{\partial t^2}=\sum_{\beta} \textbf{F}^{\beta\alpha}, 
\label{eqmotiontrans}
\end{equation}
and for rotation
\begin{equation}
I_b \frac{\partial^2 \textbf{w}^{\alpha}}{\partial t^2}=\sum_{\beta} \textbf{M}^{\beta\alpha}+\frac{1}{2}\sum_{\beta} (\textbf{R}^{\beta}-\textbf{R}^{\alpha})\times\textbf{F}^{\beta\alpha}, 
\label{eqmotionrot}
\end{equation}
where the summation of forces $\textbf{F}^{\beta\alpha}$ and torques $\textbf{M}^{\beta\alpha}$ take place over all the beads $\beta$ at the position $\textbf{R}^{\beta}$ in contact with the bead $\alpha$ of mass $m_b$, moment of inertia $I_b$, and at the position $\textbf{R}^{\alpha}$ \textcolor{black}{\cite{merkel2010PRE,merkel2017ijss,supp}. 
The FCC structure is formed by spheres of radius $r_b=5$~mm made of stainless steel with a Young modulus $E=200$~GPa, Poisson's ratio $\nu=0.3$ and density $\rho_b=7.7\cdot10^{3}$~kg.m$^{-3}$. The lattice constant of the FCC structure is $a=(2\sqrt{2})r_b$. We assume that the contacts between the particles are formed by a solid bridge with a radius $r_s=r_b/20$ and a length $h_s=r_b/50$. The different stiffnesses are $K_N = \pi E r_s^2/h_s = 3.9\cdot 10^{8}$~N/m, $K_S = K_N/(2+2\nu) = 1.5\cdot 10^{8}$~N/m, $G_T = K_S r_s^2/2 = 4.7$~Nm/rad and $G_B = K_N r_s^2/4 = 6.1$~Nm/rad \cite{brendel2011}.}
The Bloch bands structure shown in Fig.~\ref{figband}(b) depict the propagation of a longitudinal (L), rotational (R), two Transverse-Rotational (TR) and two Rotational-Transverse (RT) modes \cite{merkel2011prl}. The influence of the rotational degrees of freedom has been experimentally confirmed in granular assembly with both millimeter-sized spheres \cite{merkel2011prl} and micron-sized spheres \cite{hiraiwa2016prl} showing the wide range of frequencies and size scales where this model remains valid. The RT and R modes are optical-type modes and have no equivalent in continuum elastic solids.
Thanks to the existence of these two type of modes, the bulk bands display accidental degeneracies as marked by squares, triangles, diamonds and hexagons in Fig.~\ref{figband}(b) and Fig.~\ref{figband}(d) in the highlighted zone around 150 kHz. These degeneracies are accidental because of their dependence on the values of the stiffnesses. Owing to the symmetries of the lattice that prohibit a gap opening, the band crossings are not discrete in the reciprocal space but extend to nodal rings in the Brillouin zone, as shown in Fig.~\ref{figband}(d). The formation of the intersecting nodal rings stems from three types of symmetries, namely the mirror reflection symmetry, the time-reversal along with the space inversion symmetries and the nonsymmorphic group with glide plane or screw axes symmetry \cite{ruiyu2017prl}. On account of the $O_h$ point group symmetry of the FCC lattice, the (001) plane, comprising either the $k_x$, $k_y$ or $k_z=0$ planes, is a mirror reflection plane and thus hosts here two overlaying nodal rings with frequencies ranging from 148 to 152 kHz. In the frequency range spanning from 147 to 157 kHz, two nodal rings lie in and are protected by the (110) mirror plane, here shown as the $k_y=k_z$ plane. The high-symmetry direction $\Gamma$--X joins the (110) and (001) surfaces, therefore the aforementioned inner nodal chains are interconnected in the form of outer nodal chains at the point marked with an orange square as shown in Fig.~\ref{figband}(d) \cite{PhysRevLett.119.156401}. Similarly, the high symmetry direction $\Gamma$--Z also connects these two symmetry planes, thus, these two nodal chains merge again at the point marked with a triangle. 
The (111) plane contains the space inversion symmetry along with the time reversal symmetry that is always preserved in our case, consequently, two concentric nodal rings are protected across this surface, however, slightly off-plane \cite{hirayama2017}. Since the bands crossings that are marked with orange diamonds and hexagons occur across the L--U and L--K directions that extend onto the (110) plane, the four nodal rings intersect in an interlaced combination of outer nodal chains and a Hopf link.

\begin{figure}[htbp]
\begin{center}
\includegraphics[width=0.6\columnwidth]{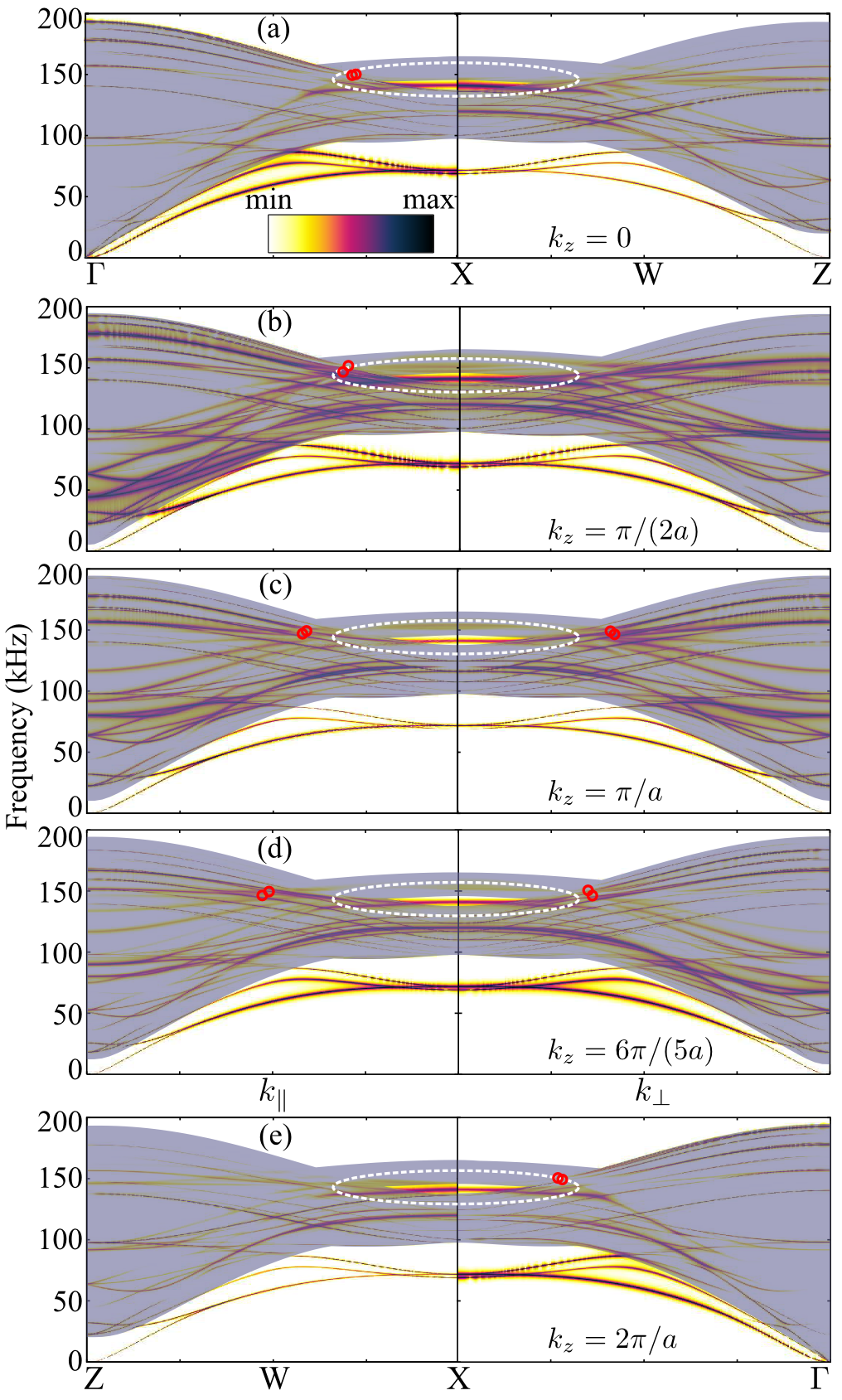}
\caption{Topological surface states for the granular metamaterial on the (001) interface. The gray shaded area depict the projected bulk bands. $k_{\parallel}$ is oriented along the $\Gamma$--X direction whereas $k_{\perp}$ is oriented along the X--Z direction. The white dashed ellipses highlight the topological surface states. The red circles mark the positions of the nodal lines as reported in Fig.~\ref{figband}(d). \label{figsurface}}
\end{center}
\end{figure}
We now investigate the existence of topologically protected surface states on the (001) plane by means of numerical simulations in a finite FCC granular metamaterial slab, which remains periodic only in the $x$ and $y$ directions but has stress-free boundary conditions at the upper and lower crystal terminations. 
\textcolor{black}{The numerical simulations are carried out using a Discrete Particle Method with the code MercuryDPM \cite{mercury1,mercury2,mercury3} and the dispersion bands are found by computing the two-dimensional Fourier transform in time and space along one unique direction \cite{mouraille2006,merkel2017ijss,supp}. }
In order to unravel these geometrically protected surface excitations in the entire 3D space, we compute them along the usual in-plane components of the wavevector but evaluate the band diagram for various out-of-plane wavevector components $k_z$. Technically speaking, we implemented this by an emitting bead-array whose phase difference accounts for different wave propagation momenta. The band diagram of the granular metamaterial slab, which is presented in Fig.~\ref{figsurface} is characterized by red circles, also seen in Fig.~\ref{figband}(d), indicating the dispersion along the rim of the nodal ring at $k_y=0$. The gray shaded background of the diagrams illustrate the projected bulk bands sustained by the slab containing the stress-free surfaces along the $z$ direction. With seemingly little dispersion with varying out-of-plane momentum $k_z$, Fig.~\ref{figsurface} illustrates the presence of a surface state within the incomplete band gaps that are confined to the nodal ring. Specifically, as rendered by the white dashed area, the bulk band gap spans from 138 to 146~kHz in between which, an entirely flat surface state resides (141~kHz). Further, this surface excitation that transpires from the degenerate bulk states (red circles) with a near zero group velocity, emerges from the non trivial Berry phase of the nodal rings. In addition, as expected, at lower frequencies trivial surface states that are associated to the stress-free boundary condition are also excited as seen in Fig.~\ref{figsurface}. 

\begin{figure}[htbp]
\begin{center}
\includegraphics[width=0.8\columnwidth]{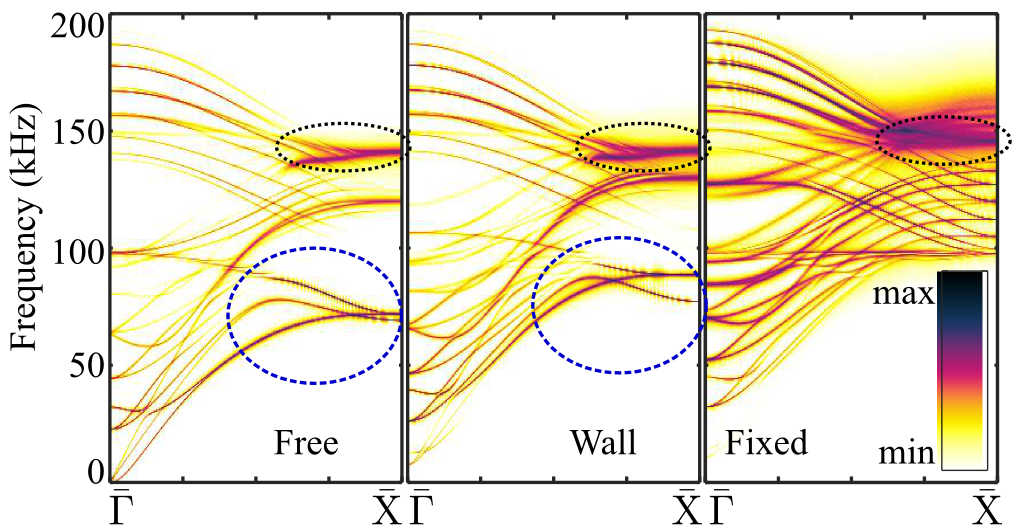}
\caption{Ultrasonic band diagram along the $\bar{\Gamma}$--$\bar{\text{X}}$ direction of topological surface states in a finite slab. We consider three different crystal terminations: free, in contact with a wall, and fixed. The blue dashed (black dotted) circle mark the spectral region of the topologically trivial (non-trivial) surface states. \label{figboundary}}
\end{center}
\end{figure}

In order to shed light on the topological properties of the surface waves discussed above, we compare three different crystal terminations. In addition to the free boundary condition, we consider the cases where the top layer is in contact with a plane rigid wall and where the top layer is in contact with a layer of fixed (immovable) particles placed in the continuity of the crystalline structure. These three crystal terminations produce an interface between the granular metamaterial and a medium into which sound waves cannot leak. A single chain of excited particles parallel to the $y$ axis acts as the source, hence, all kind of states can be easily excited parallel to the crystal termination for all possible values of $k_z$. In the numerical experiments, we probe the mechanical response to the excitation in the nearest vicinity of the termination. As one can see in Fig. \ref{figboundary} within the dashed blue ellipses, the trivial surface states at lower frequencies are highly sensitive to the specific boundary condition employed. Beyond significant shifts and enforced dispersion, the trivial surface state ceases to exist in the presence of a fixed crystal termination. On the contrary, the surface state emerging from the nodal rings remains intact and within the bulk band gap in every scenario as highlighted by the black dotted ellipses, underlining its topological origin and resilience against drastic interface perturbations.

\begin{figure}[H]
\begin{center}
\includegraphics[width=0.6\columnwidth]{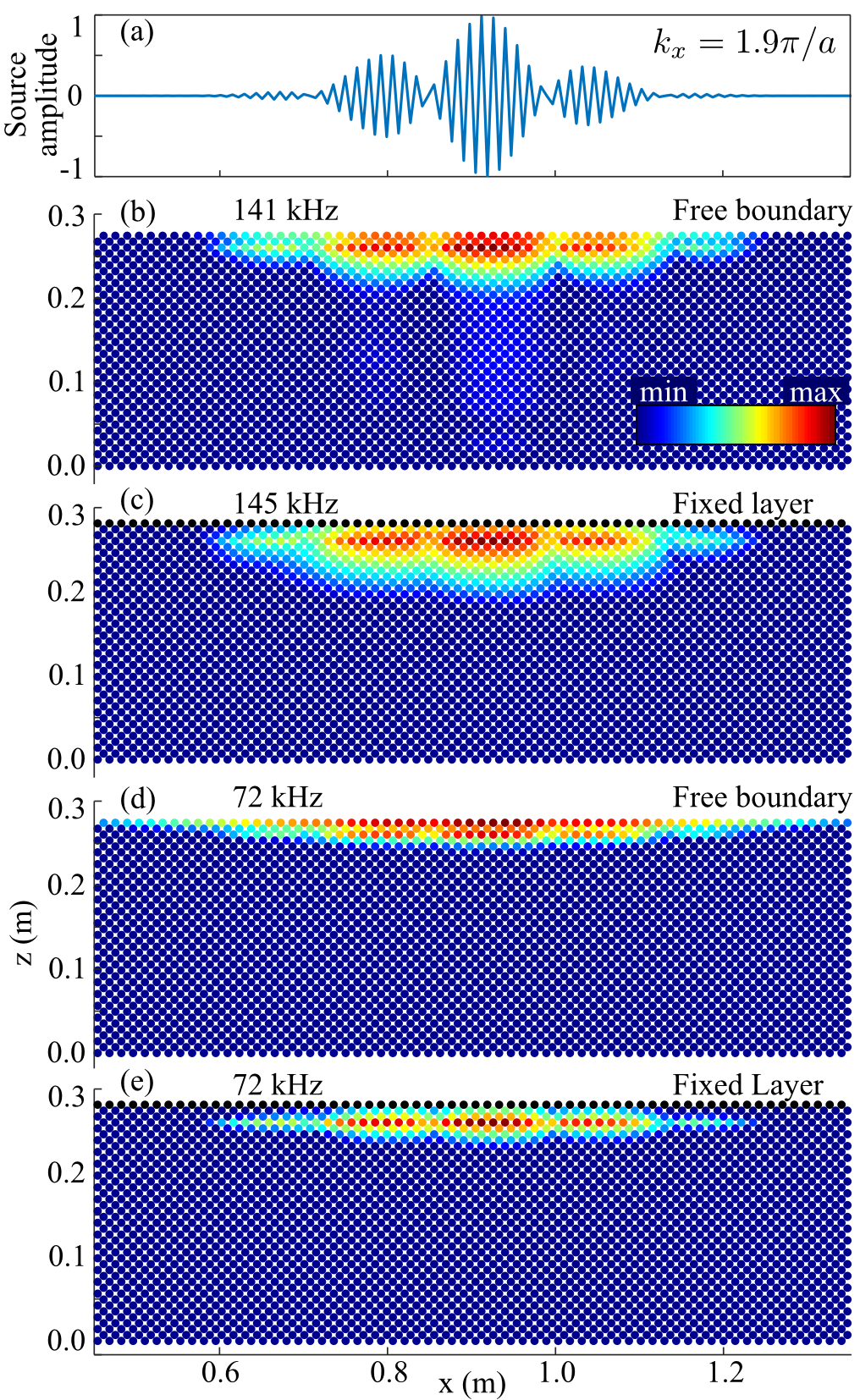}
\caption{Launching of surface states parallel to the terminated crystal interface (001) (along the $\Gamma$--X direction) by placing a harmonic source at the ante-penultimate layer. (a) Spatial source profile to momentum match, $k_x=1.9\pi/a$, the trivial and non-trivial surface states at their corresponding frequencies. (b,c) Non-trivial surface state with a stress free termination and a fixed (clamped) termination. (d) Conventional or trivial surface wave with a stress free termination. (e) Fixed termination, here no trivial surface state exists.  \label{figsurfacemode}}
\end{center}																			
\end{figure}

The above mentioned robustness against interface perturbation of the topological surface states emanating the nodal rings is visualized by computing the mechanical motions as seen in Fig.~\ref{figsurfacemode}. At the ante-penultimate layer we place a source array with a predefined spatial phase profile, as rendered in Fig.~\ref{figsurfacemode}(a), to momentum match ($k_x=1.9\pi/a$, where $a$ is the lattice constant) the two distinct surface states computed in Fig. \ref{figboundary} at their corresponding frequencies. In Fig.~\ref{figsurfacemode}(b) and Fig.~\ref{figsurfacemode}(d) we compute the spatial mechanical field of the surface states confined along the stress free crystal termination. The topological non-trivial (trivial) state at 141~kHz (72~kHz) displays a strong field confinement, it appears however that the low frequency excitation displays a smaller penetration depth. Finally, we introduce a fixed termination above the top layer. Apart from an unremarkable frequency shift, the mode shape of the confined topological non-trivial surface state at 145~kHz persists virtually unaffected when comparing Fig.~\ref{figsurfacemode}(b) and Fig.~\ref{figsurfacemode}(c). With a fixed termination, as computed in Fig. \ref{figboundary}, the trivial state ceases to exist, hence at 72~kHz the field remains solely localized around the excitation source as illustrated in Fig. \ref{figsurfacemode}(e). \\
In conclusion, three-dimensional man-made FCC granular metamaterials have been found to host complex structures of nodal rings interlaced in the form of inner and outer chains. The symmetry protected nodal features give rise to remarkably robust surface states that appear entirely unaffected by the type of crystal termination. Hence, our results provide a convincing platform to implement topological non-trivial nodal rings and chains physics for mechanical waves, potentially useful for robust ultrasonic sensing and transduction.

\section{Acknowledgements}
J. C. acknowledges the support from the European Research Council (ERC) through the Starting Grant No. 714577 PHONOMETA and from the MINECO through a Ram\'on y Cajal grant (Grant No. RYC-2015-17156).

\newpage
%

\end{document}